\documentstyle[aps,multicol,epsf]{revtex}

\newcommand{\be}{\begin{equation}}
\newcommand{\ee}{\end{equation}}
\newcommand{\bea}{\begin{eqnarray}}
\newcommand{\eea}{\end{eqnarray}}
\newcommand{\ba}{\begin{array}}
\newcommand{\ea}{\end{array}}

\begin{document}
\title{
  \begin{flushright} \begin{small}
   hep-th/0004096
  \end{small} \end{flushright}
\vspace{1.cm}
Bianchi Type I Cosmologies in Arbitrary Dimensional Dilaton Gravities}
\author{Chiang-Mei Chen\footnote{E-mail: cmchen@joule.phy.ncu.edu.tw}}
\address{Department of Physics, National Central University,
         Chungli 320, Taiwan}
\author{T. Harko\footnote{E-mail: tcharko@hkusua.hku.hk}
    and M. K. Mak\footnote{E-mail: mkmak@vtc.edu.hk}}
\address{Department of Physics, University of Hong Kong,
         Pokfulam, Hong Kong}
\date{April 10, 2000}
\maketitle

\begin{abstract}
We study the low energy string effective action with an exponential type
dilaton potential and vanishing torsion in a Bianchi type I space-time
geometry. In the
Einstein and string frames the general solution of the gravitational
field equations can be expressed in an exact parametric form.
Depending on the values of some parameters the obtained cosmological
models can be generically divided into three classes, leading to both
singular and nonsingular behaviors.
The effect of the potential on the time evolution of the mean anisotropy
parameter is also considered in detail, and it is shown that a Bianchi
type I Universe isotropizes only in the presence of a dilaton field
potential or a central deficit charge.
\end{abstract}
\vspace{.5cm}
\hspace{1cm}{PACS number(s): 04.20.Jb, 04.65.+e, 98.80.-k}

\begin{multicols}{2}
\narrowtext
\section{Introduction}
In an attempt to address the potential, inherited from string theory, to
eliminate the initial cosmological singularity, from which {\em time} and
our Universe are supposed to have begun about 15 billion years ago,
Gasperini and Veneziano initiated a program known as the
pre-big bang scenario \cite{GaVe93}.
The field equations of the pre-big bang cosmology are based on
the low energy effective action resulting from string theory.
In $D$-dimensions, the massless bosonic fields from the NS-NS sector
are the dilaton, $\phi$, the antisymmetric tensor, $B_{\mu\nu}$, and
the metric tensor, $\hat g_{\mu\nu}$, whose dynamics
is described, in the ``string frame'', by the following action:
\be
\hat S = \int d^Dx \sqrt{-\hat g} e^{-2\phi} \left\{
     \hat R + \hat \kappa (\hat \nabla \phi)^2 - \frac1{12} H_{[3]}^2
   - \hat U(\phi) \right\}, \label{hS}
\ee
where $H_{[3]}=d B_{[2]}$ and $\hat \kappa$ is a generalized dilaton
coupling constant ($\hat \kappa=4$ for superstring theories).
Moreover, we also allow for the existence of a potential $\hat U(\phi)$
of the dilaton field.
From a physical point of view the most important candidate for the
potential is a cosmological constant $\Lambda$, which appears
in the massive extension of type IIA supergravity
and is restricted to be positive, $\Lambda>0$ \cite{Ro86}.

For simplicity in the following we shall assume that $H_{[3]}$ is
vanishing. In this circumstance, via a conformal rescaling
\be
g_{\mu\nu} = e^{-\frac4{D-2}\phi} \hat g_{\mu\nu}, \label{ghatg}
\ee
the action (\ref{hS}) reduces to a $D$-dimensional dilaton gravity whose
action, in the ``Einstein frame'', has the form
\be
S = \int d^Dx \sqrt{-g} \left\{ R
    - \kappa(\nabla \phi)^2 - U(\phi) \right\},
    \label{S}
\ee
with $U(\phi) = e^{\frac{4\phi}{D-2}} \hat U(\phi)$ and
$\kappa=\frac{4(D-1)}{D-2} - \hat\kappa$.

Pre-big bang inflationary cosmological models, based on the actions
(\ref{hS}) or (\ref{S}) have been recently intensively investigated
in the physical literature
\cite{GaRi95,BrEaMa98,MaSh97,EaMaWa96,DiDe99,LuOvWa98,Ga99,CoLaWa94}.
Gasperini and Ricci \cite{GaRi95} have obtained exact solutions to the
four-dimensional low energy string effective action adopting a
space-independent dilaton and vanishing Kalb-Ramond anti-symmetric
tensor fields ans\"atz for the Bianchi type
I, II, III, V, VI${}_0$ and VI${}_h$ geometries.
They have shown that in such a context the initial curvature
singularities can not be avoided.
Brandenberger, Easther and Maia \cite{BrEaMa98} have found non-singular
spatially homogeneous and isotropic solutions for dilaton gravity in the
presence of a special combination of higher derivative terms in the
gravitational action.
Some of these solutions correspond to a spatially flat, bouncing
Universe originating in a dilaton-dominated contracting phase and
emerging as an expanding FRW Universe.

Very recently, the string cosmology equations with a dilaton potential
have been examined, in the string frame, by Ellis {\sl et al.}
\cite{ElRoSoDu99},
who also give a generic algorithm for obtaining solutions with desired
evolutionary properties.
The presence of a dilaton potential leads to the violation of the
pre-big bang symmetry $a(t) \to 1/a(t)$.
Moreover, Garcia de Andrade \cite{GaA99} obtained several classes of
solutions of the Einstein-Cartan dilatonic inflationary cosmology.
In the cases where the dilatons are constrained by the presence of
spin-torsion effects a repulsive gravity is found.
The temperature fluctuation has also been computed from the nearly flat
spectrum of the gravitational waves produced during inflation, with
results agreeing with the COBE data.

Pre-big bang cosmological models, in which there is no need to introduce
the inflation or to fine-tune potentials, have many attractive features
\cite{Ve98}.
Inflation is natural, thanks to the duality symmetries of string cosmology,
and the initial condition problem is decoupled from the singularity problem.
Finally, quantum instability (pair creation) is able to heat up an
initially cold Universe and generate a standard hot big bang with the
additional features of homogeneity, flatness and isotropy.

It is the purpose of the present paper to study Bianchi type I cosmological
models in the dilaton gravity (\ref{hS}) and (\ref{S}).
More specifically, we shall consider the effects of an exponential type
potential, $U(\phi)=U_0 e^{\lambda\phi}$, with arbitrary values of the
constants $U_0, \lambda$, on the dynamics and evolution of an anisotropic
space-time, in both the Einstein and string frames.
In this case the general solution of the gravitational
field equations can be expressed in an exact parametric form.
The physical effects of the potential on the evolution of the anisotropic
space-time are also considered in detail.

The present paper is organized as follows. The basic equations describing
the dilatonic Bianchi type I cosmological model are obtained
in Section 2. The general solution of the field
equations for an exponential type dilaton potential is obtained
in Section 3 (Einstein frame) and in Section 4 (string frame).
In Section 5 we discuss our results and conclusions.

\section{Einstein Frame Field Equations, Geometry and Consequences}
In this paper, we shall consider the $D$-dimensional anisotropic
generalization of the flat FRW geometry --- the Bianchi type I space-time
described by the line-element
\be
ds^2 = -dt^2 + \sum_{i=1}^{D-1} a_i^2(t) (dx^i)^2.
\ee
For this metric, it is convenient to introduce the following variables:
{\em volume scale factor} $V$, {\em directional Hubble factors} $H_i$
and {\em mean Hubble factor} $H$ as
\bea
V &:=& \prod_{i=1}^{D-1} a_i, \label{V} \\
H_i &:=& \frac{\dot a_i}{a_i}, \quad i=1,...,D-1, \\
H &:=& \frac1{D-1} \sum_{i=1}^{D-1} H_i, \\
\Delta H_i &:=& H_i - H, \quad i=1,...,D-1. \label{DeltaH}
\eea
Then one can immediately check out the relation
\be
H = \frac1{D-1} \frac{\dot V}{V}.
\ee
In terms of variables (\ref{V})-(\ref{DeltaH}) the Ricci tensor
of the Bianchi type I geometry can be expressed as
\bea
R_{00} &=& -\frac{d}{dt}
      \sum_{i=1}^{D-1} \left( \frac{\dot a_i}{a_i} \right)
    - \sum_{i=1}^{D-1} \left( \frac{\dot a_i}{a_i} \right)^2 \nonumber\\
    &=& -(D-1) \dot H - \sum_{i=1}^{D-1} H_i^2, \\
R_{ii} &=& a_i^2 \left[ \dot H_i + (D-1)H H_i \right], \quad i=1,...,D-1.
\eea

On the other hand, the field equations of the action (\ref{S}) can be
achieved by variation with respect to the fields $g^{\mu\nu}$ and $\phi$
giving
\bea
R_{\mu\nu} - \kappa \nabla_\mu \phi \nabla_\nu \phi
    - \frac{U}{D-2} g_{\mu\nu} &=& 0, \\
\nabla^2 \phi - \frac1{2\kappa} \frac{\partial U}{\partial \phi} &=& 0,
\eea
where $\nabla$ is the covariant derivative of $g_{\mu\nu}$.
Thus, for the Bianchi type I space-time, the gravitational field equations
in the Einstein frame reduce to
\be
(D-1) \dot H + \sum_{i=1}^{D-1} H_i^2 + \kappa \dot\phi^2
    - \frac1{D-2} U = 0, \label{EqH}
\ee
\bea
\frac1{V} \frac{d}{dt} (VH_i) - \frac1{D-2} U &=& 0, \quad i=1,..,D-1,
      \label{EqV} \\
\frac1{V} \frac{d}{dt} (V \dot\phi) + \frac1{2\kappa}
      \frac{\partial U}{\partial\phi} &=& 0. \label{Eqphi}
\eea

By summing equations (\ref{EqV}) we obtain
\be
\frac1{V} \frac{d}{dt} (VH) = \frac1{D-2} U, \label{EqHV}
\ee
which, together with (\ref{EqV}), leads to
\be
H_i = H + \frac{K_i}{V}, \quad i=1,...,D-1. \label{Hi}
\ee
In equations (\ref{Hi}) $K_i, i=1,...,D-1$ are constants of integration,
which satisfy the relation:
\be
\sum_{i=1}^{D-1} K_i = 0.
\ee

Substituting eqs.(\ref{Hi}) into (\ref{EqH}) and then combining with
eq.(\ref{EqHV}) we obtain
\be
\kappa \dot \phi^2 + (D-2) \dot H + \frac{K^2}{V^2} = 0, \label{kdp}
\ee
where $K^2:=\sum_{i=1}^{D-1} K_i^2$.
Consequently, the remaining task is to solve the equations (\ref{Eqphi}),
(\ref{EqHV}) and (\ref{kdp}).

The physical quantities of interest in cosmology are
the {\em expansion scalar} $\theta$,
the {\em mean anisotropy parameter} $A$,
the {\em shear scalar} $\Sigma^2$ and
the {\em deceleration parameter} $q$ defined according to:
\bea
\theta &:=& (D-1) H = \frac{\dot V}{V}, \\
A &:=& \frac1{D-1}\sum_{i=1}^{D-1} \left( \frac{\Delta H_i}{H} \right)^2
    = \frac1{D-1} \frac{K^2}{V^2 H^2}, \\
\Sigma^2 &:=& \frac1{D-2} \left( \sum_{i=1}^{D-1} H_i^2-(D-1)H^2 \right)
       \nonumber\\
   &=& \frac{D-1}{D-2} AH^2, \\
q &:=& \frac{d}{dt} H^{-1} - 1.
\eea
The sign of the deceleration parameter indicates whether the
cosmological model inflates. The positive sign corresponds to standard
decelerating models whereas the negative sign indicates inflationary
behavior.

\section{Exponential Potential in the Einstein Frame}
The cosmological behavior of Universes filled with scalar field, $\phi$,
as well as a Liouville type exponential potential
\be
U(\phi) = U_0 e^{\lambda \phi}, \label{expU}
\ee
with $U_0$ and $\lambda$ constants, has been extensively investigated in
the physical literature for both homogeneous and inhomogeneous scalar
fields \cite{Ba87}-\cite{BySc98}.
An exponential potential arises in the four-dimensional effective
Kaluza-Klein type theories from compactification of the higher-dimensional
supergravity or superstring theories \cite{CaMaPeFr85}.
A solution in the case of a flat space-time filled with a scalar field
with an exponential potential but describing power-law inflationary
behavior has been obtained by Barrow \cite{Ba87}.
Higher dimensional ($D\ge 4$) anisotropic cosmological models with a
massless scalar field self-interacting through an exponential potential
have been investigated in \cite{AgFeIb93a}.
A non-inflationary solution for an open FRW Universe
exponential-potential pure scalar field filled space-time and with
scalar field energy density decaying as $\rho_\phi \sim t^{-2}$ has been
recently found by Mubarak and Oezer \cite{MuOe98}.
In the Einstein frame the exponential potential (\ref{expU}) is also
generated by means of the conformal transformation (\ref{ghatg})
for $\hat U(\phi)=\Lambda$, with $\Lambda$ the central charge deficit.

For this type of potential, the combination of equations (\ref{EqV}) and
(\ref{Eqphi}) leads to
\be
\frac{d}{dt} \left( \frac1{D-2} V \dot \phi
    + \frac{\lambda}{2\kappa} VH \right) = 0,
\ee
or, equivalently, to
\be
\dot \phi = \frac{(D-2)C}{V} - \frac{(D-2) \lambda}{2\kappa} H, \label{C0}
\ee
with $C$ a constant of integration.

Substitution of Eq.(\ref{C0}) into Eq.(\ref{kdp}) gives the ``final''
field equation
\be
\frac{\ddot V}{V} + \alpha \frac{\dot V^2}{V^2} - \beta \frac{\dot V}{V^2}
    + \gamma \frac1{V^2} = 0, \label{alpha}
\ee
where
\bea
\alpha &=& \frac{(D-2)\lambda^2}{4(D-1)\kappa} - 1, \\
\beta &=& (D-2) C \lambda, \\
\gamma &=& (D-1)(D-2) C^2 \kappa + \frac{(D-1)K^2}{D-2}.
\eea

By introducing a new variable $u := \dot V$,
equation (\ref{alpha}) takes the form
\be
\frac{u du}{-\alpha u^2 + \beta u - \gamma} = \frac{dV}{V}. \label{Equ}
\ee
Equation (\ref{Equ}) has the general solution (with $V_0$ a constant of
integration):
\be
V = V_0 \exp \left( \int \frac{u du}{-\alpha u^2 + \beta u - \gamma}
    \right). \label{VVV}
\ee

In the following we shall denote
\bea
\Delta &=& \beta^2 - 4\alpha\gamma, \\
u_0 &=& \frac{\beta}{2\alpha}, \\
u_{\pm} &=& \frac{\beta \pm \sqrt{\Delta}}{2\alpha}, \\
m_{\pm} &=& -\frac1{2\alpha}\left(1\pm\frac{\beta}{\sqrt{\Delta}}\right).
\eea
Hence, taking $u$ as a parameter, we obtain three classes of solutions of
the gravitational field equations describing a dilaton field
filled Bianchi type I pre-big bang Universe.
The explicit form of the solutions depends on the values of the parameters
$\alpha, \beta$ and $\gamma$. All the solutions are expressed in a
closed parametric form and are given by:

\subsection{$\Delta>0$}
\bea
t &=& t_0 + V_0 \int(u-u_+)^{m_+-1}(u-u_-)^{m_--1} du, \\
V &=& V_0 (u-u_+)^{m_+} (u-u_-)^{m_-}, \\
a_i &=& a_{i0} \prod_{\epsilon=\pm}
    u^{-\frac{K_i m_\epsilon}{u_\epsilon}}
    (u-u_\epsilon)^{(\frac1{D-1}+\frac{K_i}{u_\epsilon})m_\epsilon}, \\
q &=& (D-2)+\frac{(D-1)\alpha}{u^2} \prod_{\epsilon=\pm}(u-u_\epsilon), \\
U &=& -\frac{(D-2)\alpha}{(D-1)V_0^2} \prod_{\epsilon=\pm}
    (u-u_\epsilon)^{1-2 m_\epsilon}.
\eea

\subsection{$\Delta=0$}
\bea
t &=& t_0 - \frac{V_0}{\alpha} \int (u-u_0)^{-\frac1{\alpha}-2}
      \exp \left( \frac{u_0}{\alpha(u-u_0)} \right) du, \\
V &=& V_0 (u-u_0)^{-\frac1{\alpha}}
      \exp \left( \frac{u_0}{\alpha(u-u_0)} \right), \\
a_i &=& a_{i0} (u-u_0)^{-\frac1{(D-1)\alpha}}
      \exp \left( \frac{u_0+K_i(D-1)}{\alpha(D-1)(u-u_0)} \right), \\
q &=& (D-2) + \frac{(D-1)\alpha(u-u_0)^2}{u^2}, \\
U &=& -\frac{(D-2)\alpha}{(D-1)V_0^2} (u-u_0)^{\frac2{\alpha}+2}
      \exp \left( \frac{-2u_0}{\alpha(u-u_0)} \right).
\eea

\subsection{$\Delta<0$}
\bea
t &=& t_0 + V_0 \int (-\alpha u^2+\beta u-\gamma)^{-\frac1{2\alpha}-1}
    \nonumber\\
  && \times \exp \left( -\frac{\beta}{\alpha\sqrt{-\Delta}}
      \arctan \frac{2\alpha u-\beta}{\sqrt{-\Delta}} \right) du, \\
V &=& V_0 (-\alpha u^2+\beta u-\gamma)^{-\frac1{2\alpha}} \nonumber\\
  && \times \exp \left( -\frac{\beta}{\alpha\sqrt{-\Delta}}
      \arctan \frac{2\alpha u-\beta}{\sqrt{-\Delta}} \right), \\
a_i &=& a_{i0} (-\alpha u^2+\beta u-\gamma)^{-\frac1{2\alpha(D-1)}}
    \nonumber\\
  && \times \exp \left( -\frac{\beta+2\alpha K_i(D\!-\!1)}
                              {\alpha(D-1)\sqrt{-\Delta}}
      \arctan \frac{2\alpha u-\beta}{\sqrt{-\Delta}} \right), \\
q &=& (D-2)-(D-1)\frac{-\alpha u^2+\beta u-\gamma}{u^2}, \\
U &=& \frac{D-2}{(D-1)V_0^2} (-\alpha u^2+\beta u-\gamma)^{\frac1{\alpha}+1}
    \nonumber\\
  && \times \exp \left( \frac{2\beta}{\alpha\sqrt{-\Delta}}
      \arctan \frac{2\alpha u-\beta}{\sqrt{-\Delta}} \right).
\eea

For all three cases, the quantities $\theta, A$ and $\Sigma^2$ can be
easily found from
\be
\theta = \frac{u}{V}, \quad
A = \frac{(D-1)K^2}{u^2}, \quad
\Sigma^2 = \frac{K^2}{(D-2)V^2}.
\ee

\section{Exponential Potential in the String Frame}
In the string frame the gravitational field equations and the dilaton
equations are obtained by varying the action (\ref{hS}) and, under the
assumption of vanishing $H_{[3]}$, are given by
\bea
\hat R_{\mu\nu} - \frac12 \hat g_{\mu\nu} \hat R
   + 2 \hat\nabla_\mu \hat\nabla_\nu \phi
   + (\hat\kappa-4) \hat\nabla_\mu \phi \hat\nabla_\nu \phi &&\nonumber\\
   - \frac12 \hat g_{\mu\nu} \left\{ 4 \hat\nabla^2 \phi
   + (\hat\kappa-8) (\hat\nabla \phi)^2 - \hat U \right\} &=& 0,
     \label{*} \\
\hat R + \hat \kappa \hat \nabla^2 \phi - \hat \kappa (\hat\nabla \phi)^2
   - \hat U(\phi) + \frac12 \frac{\partial \hat U}{\partial \phi} &=& 0.
     \label{**}
\eea
By eliminating $\hat R$ between equations (\ref{*}) and (\ref{**}), the
gravitational field and dilaton equations take the form
\bea
\hat R_{\mu\nu} + 2 \hat\nabla_\mu \hat\nabla_\nu \phi
   + (\hat\kappa-4) \hat\nabla_\mu \phi \hat\nabla_\nu \phi
   \qquad\qquad && \nonumber\\
   - \frac{\hat g_{\mu\nu}}2 \left\{ (4\!-\!\hat\kappa) \hat\nabla^2 \phi
   + 2 (\hat\kappa\!-\!4) (\hat\nabla \phi)^2
   - \frac12 \frac{\partial \hat U}{\partial \phi} \right\} &=& 0, \\
\hat\nabla^2 \phi - 2(\hat\nabla \phi)^2
   + \frac{4 \hat U + (D-2) \frac{\partial \hat U}{\partial \phi}}
          {2[(D-2)\hat\kappa-4(D-1)]} &=& 0.
\eea

In the present section we shall consider the
general solution of equations (\ref{*}) and (\ref{**}) for an exponential
type potential, $\hat U(\phi) = \hat U_0 \exp (\hat \lambda \phi)$, with
$\hat \lambda$ an arbitrary constant.
Since the metric tensors are connected via the conformal
transformation (\ref{ghatg}), in the string frame the general solutions
of the gravitational field equations can be obtained by applying the
conformal transformation (\ref{ghatg}) to the solution obtained in the
Einstein frame.
In the string frame we shall also assume an anisotropic Bianchi type I
geometry with line element
\be
d \hat s^2 = -d \hat t^2 + \sum_{i=1}^{D-1} \hat a_i^2(\hat t) (dx^i)^2,
\ee
with the metric tensor components in the two frames connected by the
conformal transformation (\ref{ghatg})
and with the time coordinate $\hat t$ defined according to
\be
\hat t = \int \exp \left[ \frac2{D-2} \phi(t) \right] dt.
\ee

In the two frames the volume scale factor, the directional Hubble factors
and the mean Hubble factor are related by means of the general relations:
\bea
\hat V &=& V e^{\frac{2(D-1)}{D-2} \phi} ,\\
\hat H_i &=& \left(H_i + \frac2{D-2}\dot\phi \right)e^{-\frac2{D-2}\phi},
    \quad i=1,...,D-1, \\
\hat H &=& \left(H + \frac2{D-2}\dot\phi \right)e^{-\frac2{D-2}\phi}.
\eea

To apply the conformal transformation, we need first to find the conformal
transformation factor $e^{\phi}$. From equation (\ref{EqHV}) it is easy to
obtain that the potential $U(\phi)$ can be expressed as
\be
U(\phi) = \frac{D-2}{D-1}\frac{u}{V^2}\left(\frac{d\ln V}{du}\right)^{-1},
\ee
leading to
\be
e^{\phi} = \left[ \frac{D-2}{(D-1)U_0} \right]^{\frac1{\lambda}}
    \left( \frac{V^2}{u} \frac{d\ln V}{du} \right)^{-\frac1{\lambda}}.
\ee

Therefore in the string frame the general solution of the gravitational
field equation for a dilaton field filled Bianchi type I with an
exponential potential of the form
\be
\hat U(\phi) = U_{0} \exp \left[ \left( \lambda-\frac4{D-2} \right)
     \phi \right],
\ee
with $\lambda$ an arbitrary constant, can be expressed again in an
exact closed parametric form, with $u$ taken as parameter, and is given by
\bea
\hat t - \hat t_0 &=&
\left[ \frac{D-2}{(D-1)U_{0}} \right]^{\frac2{(D-2)\lambda}} \nonumber\\
 && \times \int
\left( \frac{V^2}{u} \frac{d\ln V}{du} \right)^{1-\frac2{(D-2)\lambda}}
\frac{du}{V}, \label{hatt}
\eea
\be
\hat V = V
\left[\frac{D-2}{(D-1)U_{0}}\right]^{\frac{2(D-1)}{(D-2)\lambda}}
\left(\frac{V^2}{u} \frac{d\ln V}{du}\right)^{-\frac{2(D-1)}{(D-2)\lambda}},
\ee
\bea
\hat H &=&
\left[\frac{D-2}{(D-1)U_{0}}\right]^{-\frac2{(D-2)\lambda}}
\left(\frac{V^2}{u} \frac{d\ln V}{du}\right)^{\frac2{(D-2)\lambda}}
    \nonumber\\ && \times
\left[ \frac{2C}{V}+\frac{\kappa-\lambda}{(D-1)\kappa} \frac{u}{V} \right],
\eea
\bea
\hat a_i &=& a_{i0}
\left[\frac{D-2}{(D-1)U_{0}}\right]^{\frac2{(D-2)\lambda}}
\left(\frac{V^2}{u} \frac{d\ln V}{du}\right)^{-\frac2{(D-2)\lambda}}
 V^{\frac1{D-1}} \nonumber\\ && \; \times
\exp \left( K_i \int \frac1{u} \frac{d\ln V}{du} du \right), \;
 i=1,...,D-1,
\eea
\be
\hat A = (D-1) \sum_{i=1}^{D-1} \left[
\frac{\kappa K_i}{(\kappa-\lambda)u + 2\kappa C(D-1)} \right]^2,
\ee
\be
\hat q = (D-2)-\frac{u}{D-1} \left( \frac{d\ln V}{du}\right)^{-1}
\left[ 2C+\frac{\kappa-\lambda}{(D-1)\kappa}u\right]^{-2},
\ee
\be
\hat U = U_0\left[ \frac{D-2}{(D-1)U_{0}} \right]^{1-\frac4{(D-2)\lambda}}
     \left( \frac{V^2}{u} \frac{d\ln V}{du} \right)^{\frac4{(D-2)\lambda}-1}.
\ee

In the string frame there are also three distinct classes of solutions,
corresponding to $\Delta >0, \Delta =0$ and $\Delta <0$ respectively.
Substituting the values of $V$ obtained in the previous
section in the formulae given above, we can find, via straightforward
calculations, the explicit parametric representations, for each class of
solutions, of the general solution of the gravitational field equations
for a dilaton field filled Bianchi type
I space-time, with an arbitrary exponential potential.

If in the solution given above we take $\lambda=\frac4{D-2}$, we
obtain the general solution of the gravitational field equations in the
string frame corresponding to a constant potential, or equivalently,
to a cosmological constant.
In this case also there are three distinct classes of solutions, with
all physical quantities represented as exact functions of time.
For $\hat U(\phi) \equiv \Lambda = const.$, Eq.(\ref{hatt}) becomes
\be\label{***}
\hat t - \hat t_0 = \sqrt{\frac{D-2}{(D-1)\Lambda}} \int
  \frac{du}{\sqrt{-\alpha u^2+\beta u-\gamma}}.
\ee
In order to obtain solutions defined for all values of the parameters we
shall assume in the following that $\alpha<0$. Then Eq.(\ref{***}) has the
solutions
\bea
u &=& -\frac{\beta}{2 |\alpha |}
   + \delta_+ \cosh\frac{\hat t-\hat t_0}{\hat \tau_0},
     \quad\hbox{for}\quad \Delta > 0, \\
u &=& -\frac{\beta}{2 |\alpha |}
   + \exp \left(\frac{\hat t-\hat t_0}{\hat \tau_0} \right),
     \quad\hbox{for}\quad \Delta = 0, \\
u &=& -\frac{\beta}{2 |\alpha |}
   + \delta_- \sinh \frac{\hat t-\hat t_0}{\hat \tau_0},
     \quad\hbox{for}\quad \Delta < 0,
\eea
where we denoted $\delta_\pm:=\frac{\sqrt{\pm\Delta}}{2 |\alpha |}$
and $\hat \tau_0:=\sqrt{\frac{D-2}{(D-1)\Lambda |\alpha |}}$.

In this way we can obtain the exact (non-parametric) solution for the
anisotropic Bianchi type I geometry in the presence of a central charge
deficit. We shall not present here the resulting formulae, due to their
complicated (but elementary) mathematical form.
As compared to the Einstein frame, the evolution of the Universe in the
string frame in the presence of the cosmological constant
can be quite complicated.

\section{Discussions and Final Remarks}
In order to consider the general effects of a dilaton field potential
in the Einstein frame on the dynamics and evolution of an arbitrary
dimensional Bianchi type I space-time, we shall also give the general
solution of the gravitational field equations (\ref{EqH})-(\ref{Eqphi})
corresponding to $U(\phi)\equiv 0$. In this case we easily obtain:
\bea
V &=& V_0 t, \\
H &=& \frac1{(D-1)t}, \label{V0} \\
a_i &=& a_{i0} \; t^{p_i}, \quad i=1,...,D-1, \\
A &=& \frac{(D-1)K^2}{V_0^2} = const., \\
q &=& D-2 = const., \\
\phi &=& \phi_0 \ln t, \label{phi0}
\eea
where $\phi_0 = \phi_0' \sqrt{\frac1{\kappa}
        \left(\frac{D-2}{D-1}-\frac{K^2}{V_0^2}\right)}$,
with $\phi_0'$ a constant of integration.
The coefficients $p_i:=\frac1{D-1}+\frac{K_i}{V_0}$ satisfy the
relations $\sum_{i=1}^{D-1} p_i = 1$ and
$\sum_{i=1}^{D-1} p_i^2 = \frac1{D-1}+\frac{K^2}{V_0^2}$.
Hence in the Einstein frame the geometry of the potential free dilaton
field is of Kasner type, but with $\sum_{i=1}^{D-1} p_i^2 \ne 1$
(if we adopt the normalization $\sum_{i=1}^{D-1} p_i^2 = 1$ then we
obtain the empty Bianchi type I Universe with $\phi \equiv 0$).
The anisotropic Bianchi type I dilaton field filled Universe does not
isotropize (the mean anisotropy parameter is a constant for all times)
and its evolution is non-inflationary with $q>0$ for all $t$.

In order to analyze the general effects of the dilaton field
potential on the dilaton field filled Bianchi type I space-time in the
Einstein frame, we shall obtain first the following anisotropy equation:
\be
\frac{dA}{dt} = -\frac{2A}{H} \left( \dot H + (D-1) H^2 \right),
\ee
which can also be written in the equivalent form
\be
\frac{dA}{dt} = -\frac{2A U(\phi)}{(D-2) H},
\ee
and integrated to give
\be
A(t) = A_0 \exp \left( -\frac2{D-2} \int_{t_0}^t \frac{U(\phi)}{H} dt
   \right).  \label{At}
\ee
In equation (\ref{At}) we denoted by $A_0$ an arbitrary constant of
integration. For $U(\phi) \equiv 0$ we always have $A \equiv A_0 = const.$
If $\int_{t_0}^t \frac{U(\phi)}{H} dt$ is a monotonically increasing
positive function of time then the presence of the dilaton field potential
will lead to the fast isotropization of the Bianchi type I space-time.

In the presence of a potential the deceleration parameter
$q=-(\dot H+H^2)/H^2$ can be expressed as
\be
q = (D - 2) - \frac{U(\phi)}{(D-2)H^2}.
\ee
If $U(\phi)\equiv 0$ in the Einstein frame the evolution of the
Universe is non-inflationary, but once the condition
$U(\phi)>(D-2)^2 H^2$ is fulfilled, the dynamics of the Bianchi
type I space-times becomes inflationary.

In the present paper we have obtained the general solution of the
gravitational field equations for a Bianchi type I space-time filled
with a dilaton field with an exponential potential in both the Einstein
and string frame.
In the Einstein frame they describe generically an expanding Universe,
with $u=\dot V \ge 0$ and with properties strongly dependent on the
numerical values of the physical parameters describing the dilaton field
and its potential.
A contracting Universe with $u=\dot V < 0$ generally does not satisfy the
condition of reality of the scale factors.
The solutions of the field equations can be classified into three classes,
according to the sign of the quantity $\Delta$.
On the other hand for solutions B and C with $\Delta=0$ and $\Delta<0$
respectively, the condition $\alpha<0$ must also be imposed to ensure
the positivity of the potential and well-defined physical quantities for
all time.
In the limit of large $u$, $u\to\infty$, all three solutions have a
similar behavior.
The mean anisotropy $A$ tends in all cases to zero, indicating that an
exponential type potential leads to the isotropization of the Universe.
In the large $u$ limit the deceleration parameter behaves as
$q=(D-2)+\alpha(D-1)$.
If the condition $\alpha<-\frac{D-2}{D-1}$, or, equivalently,
$\lambda^2 < \frac{4\kappa}{D-2}$ is fulfilled, the Universe will enter
in an inflationary phase.
For values of $\alpha$ which do not satisfy this condition the evolution
of the space-time will be generally non-inflationary.
In the same limit of large $u$ the scalar field is given by
$\phi \sim \frac{\alpha+1}{\alpha} \ln u$.

For class A solutions, the Bianchi type I dilaton field filled Universe
starts in the Einstein frame from a singular state, corresponding to the
values $u=u_+$ or $u=u_-$ of the parameter.
Hence for this model a singular state with zero values of the scale
factors is unavoidable.
But for class B of solutions the evolution of the Universe is non-singular
for $u_0<0$.
In this case the scale factors are finite for all finite values of
the parameter $u$.
Alternatively, class C models are non-singular for  values of the constants
$\alpha$ and $\gamma$ such that $\alpha<0$ and $\gamma<0$.

\begin{figure}
\epsfxsize=9cm
\centerline{\epsffile{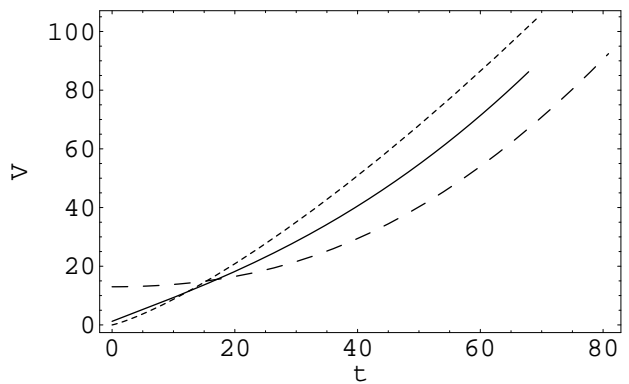}}
\caption{
  Time evolution in the Einstein frame of the four-dimensional ($D=4$)
  volume scale factor $V(t)$ of the dilaton field filled Bianchi type I
  Universe with exponential potential for different values of the
  parameters $\alpha, \beta$ and $\gamma$:
  (i). Class A Model (full curve) $\alpha=-\frac13, \beta=1, \gamma=1$,
  (ii). Class B Model (dotted curve) $u_0=1$ and
  (iii). Class C Model (dashed curve) $\alpha=-\frac13,\beta=1,\gamma=-1$.}
\label{FIG1}
\end{figure}

\begin{figure}
\epsfxsize=9cm
\centerline{\epsffile{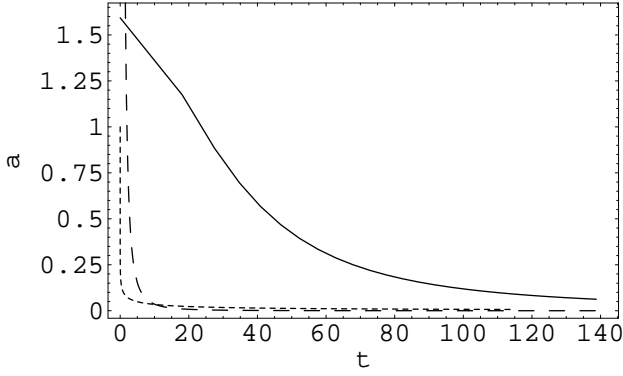}}
\caption{
  Einstein frame time evolution of the four-dimensional mean
  anisotropy parameter $a(t):=\frac{A(t)}{3K^2}$
  of the dilaton field filled Bianchi type I Universe with exponential
  potential for different values of the parameters $\alpha, \beta$ and
  $\gamma$:
  (i). Class A Model (full curve) $\alpha=-\frac13, \beta=1, \gamma=1$,
  (ii). Class B Model (dotted curve) $u_0=1$ and
  (iii). Class C Model (dashed curve) $\alpha=-\frac13,\beta=1,\gamma=-1$.
  An expanding Bianchi type I Universe always isotropizes in the presence of
  an exponential dilaton potential.}
\label{FIG2}
\end{figure}

\begin{figure}
\epsfxsize=9cm
\centerline{\epsffile{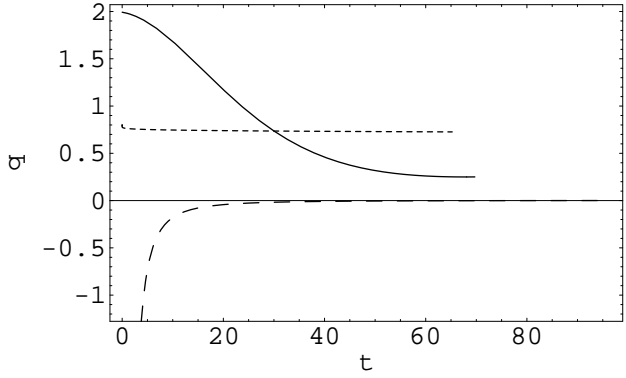}}
\caption{
  Dynamics of the four-dimensional ($D=4$) deceleration parameter $q(t)$
  of the dilaton field filled Bianchi type I Universe with exponential
  potential, in the Einstein frame, for different values of the
  parameters $\alpha, \beta$ and $\gamma$:
  (i). Class A Model (full curve) $\alpha=-\frac13, \beta=1, \gamma=1$,
  (ii). Class B Model (dotted curve) $u_0=1$ and
  (iii). Class C Model (dashed curve) $\alpha=-\frac13,\beta=1,\gamma=-1$.
  Depending on the values of the parameters the Bianchi type Universe
  has both inflationary and non-inflationary evolutions.}
\label{FIG3}
\end{figure}

\begin{figure}
\epsfxsize=9cm
\centerline{\epsffile{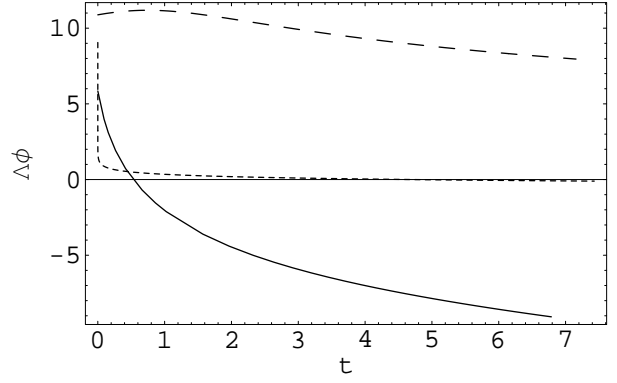}}
\caption{
  Variation in the Einstein frame of the four-dimensional ($D=4$) dilaton
  field $\phi(t)$ for different values of the parameters
  $\alpha, \beta$ and $\gamma$:
  (i). Class A Model (full curve) $\alpha=-\frac13, \beta=1, \gamma=1$,
  (ii). Class B Model (dotted curve) $u_0=1$ and
  (iii). Class C Model (dashed curve) $\alpha=-\frac13,\beta=1,\gamma=-1$.}
\label{FIG4}
\end{figure}

In Figs. \ref{FIG1}-\ref{FIG4} we have represented the variations in the
Einstein frame of the volume scale factor, mean anisotropy, deceleration
parameter and dilaton field for a four-dimensional ($D=4$) Bianchi type I
space-time.
The anisotropic Universe will always end in an isotropic state,
but its dynamics can be either inflationary or non-inflationary.
Generally the dilaton field $\phi$ is a decreasing function of time.

We shall consider now the effects of the dilaton field and potential
on the dynamics and evolution of a Bianchi type I space-time in the
string frame.
In the case in which there is no dilaton field potential,
$\hat U(\phi)\equiv 0$, the general solution of the gravitational field
equations and of the dilaton equation can be obtained again by the
conformal transformation (\ref{ghatg}) from (\ref{V0}-\ref{phi0}).
Hence in this case we obtain first the relation connecting the time
coordinate in the string and Einstein frames in the form
\be
t = \left( \frac{\hat t}{\hat n} \right)^{\hat n},
\ee
where
\be
\hat n = \frac{D-2}{D-2+2\phi_0},
\ee
In the string frame the general solution of the potential free
dilaton field filled anisotropic Universe is given by
\bea
\hat V &=& \hat V_0 \hat t^{\hat h}, \\
\hat H &=& \frac{\hat h}{(D-1)\hat t}, \\
\hat a_i &=& \hat a_{i0} \hat t^{\hat p_i}, \quad i=1,...,D-1,
\eea
and
\bea
\hat A &:=& \frac1{D-1} \sum_{i=1}^{D-1} \left(
     \frac{\Delta\hat H_i}{\hat H} \right)^2 \nonumber\\
  &=& \frac1{D-1} \sum_{i=1}^{D-1} \left[ 1
   - \frac{(D-1) \hat p_i}{\hat h} \right]^2, \\
\hat q &=& \frac{D-1}{\hat h} - 1,
\eea
where $\hat V_0$ and $\hat a_{i0}$ 
are arbitrary constants of integration.
Here we also denoted
\bea
\hat h &=& \frac{D-2+2(D-1)\phi_0}{D-2+2\phi_0}, \\
\hat p_i &=& \hat n \left(p_i + \frac{2\phi_0}{D-2}\right),
    \quad i=1,...,D-1,
\eea
and the coefficients $\hat p_i$ satisfy the relations
\bea
\sum_{i=1}^{D-1} \hat p_i
    &=& \hat n \left[ 1 + \frac{2(D-1)}{D-2}\phi_{0} \right], \\
\sum_{i=1}^{D-1} \hat p_i^2
    &=& \hat n^2 \left[ \frac1{D\!-\!1}+\frac{K^2}{V_0^2}
 + \frac{4\phi_0}{D\!-\!2} \left( 1 + \frac{D\!-\!1}{D-2} \phi_0 \right)
   \right]. \nonumber
\eea

In the string frame the general physical behavior of the potential free
dilatonic Bianchi type I Universe is quite similar to that in the Einstein
frame. The geometry is of the Kasner type, with a power-law type time
dependence of the scale factors.
The mean anisotropy of the space-time is constant and
the Universe will never isotropize. On the other hand if the condition
$\hat h > D-1$ is fulfilled, the Universe experiences an
eternal power law type inflationary anisotropic phase. Hence in the
string frame a dilaton field filled Bianchi type I Universe provides an
example of an inflating but never isotropizing cosmological type evolution.

\begin{figure}
\epsfxsize=9cm
\centerline{\epsffile{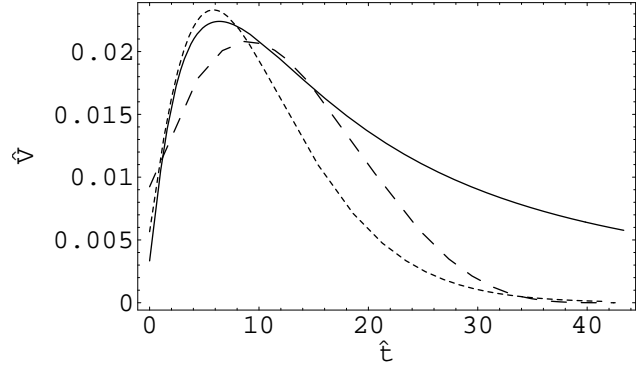}}
\caption{
  String frame evolution of the volume scale factor $\hat V$ of the dilatonic
  Bianchi type I Universe in the presence of an exponential potential
  $\hat U=U_0 \exp \left[ \left( \lambda-\frac4{D-2} \right) \phi \right]$
  as a function of time $\hat t$ for $\alpha=-1/3$, $\beta=1$, $\gamma=1$ and
  for different values of $\lambda$:
  $\lambda=3$ (full curve), $\lambda=2$ (this case corresponds to the
  presence of a central charge deficit or cosmological constant)
  (dotted curve) and $\lambda=1$ (dashed curve).
  We have used the normalization $\frac{D-2}{(D-1)U_0}=1$.}
\label{FIG5}
\end{figure}

\begin{figure}
\epsfxsize=9cm
\centerline{\epsffile{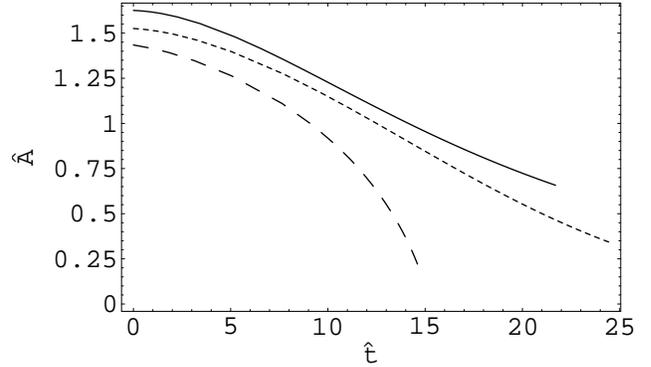}}
\caption{
  Time variation of the anisotropy parameter $\hat A$ in the string frame
  for $\alpha=-1/3$, $\beta=1$, $\gamma=1$ and for different values of
  $\lambda$:
  $\lambda=3$ (full curve), $\lambda=2$ (this case corresponds to the
  presence of a central charge deficit or cosmological constant)
  (dotted curve) and $\lambda=1$ (dashed curve).
  We have used the normalization $\frac{D-2}{(D-1)U_0}=1$.}
\label{FIG6}
\end{figure}

\begin{figure}
\epsfxsize=9cm
\centerline{\epsffile{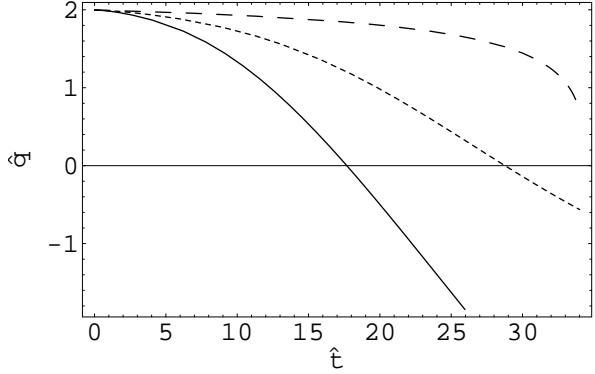}}
\caption{
  Dynamics of the deceleration parameter $\hat q$ in the string frame in
  the presence of the exponential potential
  $\hat U=U_0 \exp \left[ \left(\lambda-\frac4{D-2}\right) \phi \right]$,
  $\alpha=-1/3$, $\beta=1$, $\gamma=1$ and $\lambda=3$ (full curve),
  $\lambda=2$ (this case corresponds to the presence of a central charge
  deficit or cosmological constant) (dotted curve)
  and $\lambda=1$ (dashed curve).
  We have used the normalization $\frac{D-2}{(D-1)U_0}=1$.}
\label{FIG7}
\end{figure}

\begin{figure}
\epsfxsize=9cm
\centerline{\epsffile{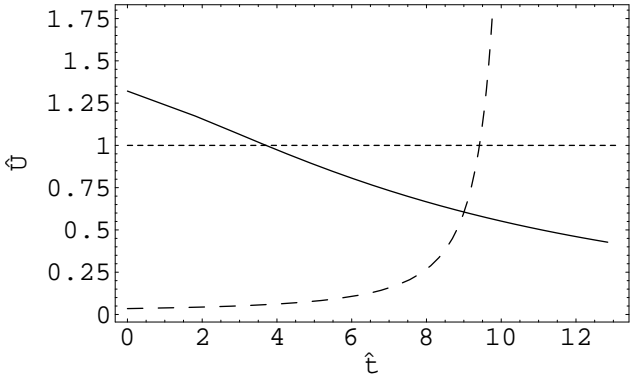}}
\caption{
  Time evolution in the string frame of the exponential potential
  $\hat U=U_0 \exp \left[ \left(\lambda-\frac4{D-2}\right) \phi \right]$
  for $\alpha=-1/3$, $\beta=1$, $\gamma=1$ and $\lambda=3$ (full curve),
  $\lambda=2$ (this case corresponds to a constant potential
  $\hat U=const.$) (dotted curve) and $\lambda=1$ (dashed curve).
  We have used the normalization $\frac{D-2}{(D-1)U_0}=1$.}
\label{FIG8}
\end{figure}

In the string frame and in the presence of an exponential potential
$\hat U(\phi)=\hat U_0 \exp \left( \hat \lambda \phi \right)$,
with $\hat \lambda =\lambda-\frac4{D-2}$ and $\lambda$ an arbitrary
constant, the Bianchi type I Universe shows a very large variety of
behaviors.
In Figs.\ref{FIG5}-\ref{FIG8} we represented the dynamics of the volume
scale factor, anisotropy parameter, deceleration parameter and potential
for different values of $\lambda$ (different exponential potential
functions) but for fixed $\alpha, \beta$ and $\gamma$.
These solutions generically begin from a singular state,
followed by an expansionary phase, with the volume
scale factor and scale factors reaching a local maximum.
Then the Universe re-collapse into a new singular phase.
This type of evolution is associated with an initial rapid isotropization
of the space-time, with the mean anisotropy parameter $\hat A$ rapidly
decreasing. Near the second
singular state the evolution of the Universe is generally inflationary,
with the string frame deceleration parameter $\hat q$ smaller than zero,
$\hat q <0$. After this phase the effect of the dilaton becomes irrelevant
to the dynamics of space-time.

In the limit of large values of the parameter $u$, the term $-\alpha u^2$
dominates, $-\alpha u^2 >> \beta u-\gamma$.
Hence in the limit of large $u$ (and large time, $\hat t \to \infty$, too),
from equation (\ref{VVV}) we obtain $V \simeq V_0 u^{-\frac1{\alpha}}$.
Therefore from Eq.(\ref{hatt}) it follows that
$\hat t \simeq u^{(1+1/\alpha)(4/(D-2)\lambda-1)} \simeq u^{1/\hat l}$,
and, consequently,
\bea
\hat V &\simeq& \hat t^{(D-1)\hat l'}, \\
\hat H &\simeq& \hat t^{-1}, \\
\hat a_i &\simeq& \hat t^{\hat l'} \exp \left(
    \frac{V_0 K_i}{\alpha} \hat t^{-\hat l} \right), \quad i=1,...,D-1,\\
\hat A &\simeq& \hat t^{-2 \hat l}, \\
\hat q &\simeq& (D-2)
   + \frac{(D-1)\alpha\kappa^2}{V_0( \kappa-\lambda )^2},\\
\hat U &\simeq& \hat t^{-2},
\eea
where we denoted
\bea
\hat l &=& \frac1{ \left( 1 + \frac1{\alpha} \right)
      \left[ \frac4{(D-2)\lambda} - 1 \right]}, \\
\hat l' &=& 1 + \hat l \left[ 1+\frac{D-2}{(D-1)\alpha} \right].
\eea

In the long-time limit the behavior of the exponential potential
dilaton field filled Universe is quite different to the behavior of
the potential free dilatonic anisotropic Universe.
The dependence of the coefficients $\hat l, \hat l'$ on the two constants
$\alpha$ and $\lambda$ leads to a larger spectrum of admissible final
states, with isotropic inflationary or non-inflationary evolution or
re-collapse into a singular state.
For $|\alpha |<1$ generally $\hat l<0$, and, if
$\hat l < - \left(1+\frac{D-2}{(D-1)\alpha}\right)^{-1}$, then the volume
scale factor tends to zero in the string frame, $\hat V \to 0$.

It is well known that the action (\ref{hS}) with vanishing antisymmetric
field strength $H_{[3]}$ is invariant with respect to scale factor duality
transformations of the form $G\to \bar G=G^{-1}$ and
$\phi \to \bar\phi-\ln(\det G)$, where $G$ is a matrix build from the
metric tensor components of the FRW, anisotropic or inhomogeneous
metric \cite{DiDe99}.
The inclusion of the potential breaks this duality,
but leads, on the other hand, to the possibility of obtaining more
general models allowing a better physical description of the very
early evolution of our Universe.

\section*{Acknowledgments}
One of the authors (TH) would like to thank Dr. P. Blaga for useful
suggestions.
CMC thanks Prof. J.M. Nester for profitable discussions.
The work of CMC was supported in part by the National Science Council
(Taiwan) under grant NSC 89-2112-M-008-016.

\end{multicols}
\end{document}